\documentclass[twocolumn,superscriptaddress,prl]{revtex4-1}
\usepackage{mathrsfs,braket}
\usepackage{amssymb, amsbsy, amsmath, latexsym, dsfont, array, layout,
graphicx,mathrsfs,color,bm}
\usepackage[normalem]{ulem}
\begin{document}

\title{An entropic effect essential for surface entrapment of bacteria}

\author{Premkumar Leishangthem}
\affiliation{Complex Systems Division, Beijing Computational Science Research Center, Beijing 100193, China}

\author{Xinliang Xu}\email{xinliang@csrc.ac.cn}
\affiliation{Complex Systems Division, Beijing Computational Science Research Center, Beijing 100193, China}
\affiliation{Department of Physics, Beijing Normal University, Beijing 100875, China}

\begin{abstract}

The entrapment of bacteria near boundary surfaces is of biological and practical importance, yet the underlying physics is still not well understood. We demonstrate that it is crucial to include a commonly neglected entropic effect arising from the spatial variation of hydrodynamic interactions, through a model that provides analytic explanation of bacterial entrapment in two dimensionless parameters: $\alpha_1$ the ratio of thermal energy to self-propulsion, and $\alpha_2$ an intrinsic shape factor. For $\alpha_1$ and $\alpha_2$ that match an {\it Escherichia coli} at room temperature, our model quantitatively reproduces existing experimental observations, including two key features that have not been previously resolved: The bacterial “nose-down” configuration, and the anticorrelation between the pitch angle and the wobbling angle. Furthermore, our model analytically predicts the existence of an entrapment zone in the parameter space defined by $\{\alpha_1,\alpha_2\}$.
\end{abstract}


\maketitle

Swimming microorganisms are constantly influenced by the presence of boundary surfaces in their natural habitat, giving rise to rich swimming behaviors \cite{Kantsler2013,Contino2015}. One phenomenon commonly known as surface entrapment attracts particular interest, where the swimmer moves along near the surface for a prolonged time \cite{Leonardo2011}. Since its first discovery in bull spermatozoa \cite{Rothchild1963}, such entrapment is widely observed for a variety of microorganisms in both domain Bacteria (e.g., Escherichia coli \cite{Bianchi2017}) and domain Eukaryota (e.g., Tetrahymena pyriformis \cite{Ohmura2018}) in different types of fluids \cite{Cao2022}, and around surfaces with different properties \cite{Poddar2020}. In addition to its importance to many biological processes such as fertilization \cite{Tung2021,Raveshi2021}, and biofilm formation \cite{Nadell2016}, such surface-swimmer interaction also provides insight for the design of microfluidic structures \cite{Dehkharghani2019} as well as artificial microswimmers \cite{Simmchen2016, Liu2016, Ketzetzi2020} for desired transport properties. However, the underlying mechanism is still poorly understood.

Using {\it E. coli} as an example where experimental data are abundant, we provide a theoretic study explaining the dynamics of flagellar bacteria near surfaces. In an early experiment, accumulation of bacteria near a solid plane was observed, which was attributed to an effective cell-plane attraction \cite{Frymier1995}. Treating one bacterium as the sum of only the leading order singularities (e.g., the force dipole, the source dipole, etc.), the hydrodynamic interaction involved in the far-field limit can be analytically obtained \cite{Berke2008, Spagnolie2012}, which predicts a seeming entrapment when extrapolated for bacteria near the wall. However, as entrapment events are typically observed at a surface distance smaller than bacterial cell size, such an extrapolation is not appropriate because all terms in the multipole expansion become important \cite{Kim2005}. At such small surface distance, the role of hydrodynamic interactions has been challenged by a demonstration where experimental data can be equally well interpreted by stochastic models with or without hydrodynamic interactions \cite{Li2009}.

Replacing the planar boundary with convex surfaces, an experiment studied the effect of surface radius, and suggested that the near field hydrodynamic interaction plays the major role \cite{Sipos2015}. In the meantime, while stochastic effects such as rotational noise \cite{Schaar2015} or bacterial tumbling \cite{Junot2022} are important for the eventual escape of bacteria at large time scales, during the entrapment they can be significantly suppressed \cite{Drescher2011, Molaei2014}. The leading role of near field hydrodynamics in surface entrapment was further confirmed by an experiment using high resolution holographic microscopy \cite{Bianchi2017}, with observation of two key features for the trapped bacteria, i.e., the “nose-down” configuration and the anticorrelation between the pitch angle and the wobbling angle. However, numerical simulations with full hydrodynamic interactions are unable to reproduce these key features: for no-slip plane the simulation results show a “nose-up” configuration \cite{Giacche2010, Pimponi2016}, and suggest a positive correlation between the wobbling angle and the pitch angle \cite{Eisenstecken2016}. In this work we show that previous theoretic/numerical studies have neglected a key entropic effect inherited from the drastic spatial variation of near field hydrodynamic interactions at room temperature. When this crucial effect is incorporated, even a simplified model of the hydrodynamic interactions is capable in providing a quantitative explanation for the observed surface entrapment behaviors.

\begin{figure}[t]
\includegraphics[width=0.5\textwidth]{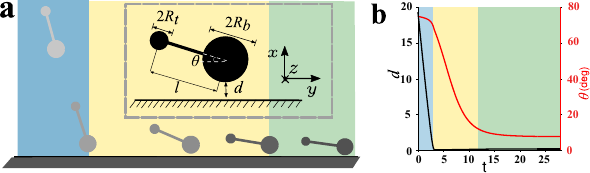}
\caption{(a) For a bacterium modeled as two spherical beads connected by a rod (inset), its trajectory in real space demonstrates surface entrapment. (b) The temporal evolutions of $d$ (black line) and $\theta$ (red line) show a three-stage dynamics visualized by blue, yellow, and green shaded areas, respectively.}\label{Fig1}
\end{figure}

\textit{Problem formulation for non-wobbling bacteria --- }For simplicity, we first study a non-wobbling bacterial model which allows analytic solution. Swimming in a fluid of viscosity $\mu$ above an infinitely large plane with no-slip boundary at $x=0$, an {\it E. coli} bacterium is simplified as two spherical beads, a body-bead with radius $R_\mathrm{b}$ and a tail-bead with radius $R_\mathrm{t}$, connected by a rigid rod that separates the two centers by $l$ (inset in Fig.\,\ref{Fig1}a) \cite{Zhang2021}. The tail-bead is propelled by a phantom force $\bm{F}_\mathrm{act}$ provided by the spinning of flagella that is not treated explicitly. Each configuration is fully determined by the surface distance $d$ between bacterial body and the plane, and pitch angle $\theta$ which is positive for a “nose-down” configuration.

Since the characteristic size and speed are about 1 $\mathrm{\mu m}$ and 10 $\mathrm{\mu m/sec}$, respectively, in water the corresponding Reynolds number is low ($10^{-5}$) so that bacterial flows are typically studied by the linear Stokes equation. At a time resolution $\Delta t\approx10^{-2} \mathrm{sec}$ the system is in the over-damped limit described by:
\begin{equation}\label{Langevin}
\bm{\mathrm{\xi}}\cdot\bm{\mathrm{U}}=\bm{\mathrm{F}}^\mathrm{P}+\bm{\mathrm{F}}^\mathrm{B},
\end{equation}
where the resistance tensor $\bm{\mathrm{\xi}}$ for any configuration is fully determined by hydrodynamics, $\bm{\mathrm{U}}\equiv\left(\bm{u}_\mathrm{b}, \bm{\omega}_\mathrm{b}, \bm{u}_\mathrm{t}, \bm{\omega}_\mathrm{t}\right)^T$ is the translational/rotational velocity vector with indices “$\mathrm{b}$" and “$\mathrm{t}$" standing for body-bead and tail-bead, respectively, $\bm{\mathrm{F}}^\mathrm{P}\equiv\left(\bm{F}_\mathrm{b}, \bm{L}_\mathrm{b}, \bm{F}_\mathrm{t}, \bm{L}_\mathrm{t}\right)^T$ represents the nonhydrodynamic forces and $\bm{\mathrm{F}}^\mathrm{B}$ represents stochastic forces.

At absolute-zero temperature, $\bm{\mathrm{F}}^\mathrm{B}=0$. Eq.\,\ref{Langevin} then reduces to $\bm{\mathrm{\xi}}\cdot\bm{\mathrm{U}}=\bm{\mathrm{F}}^\mathrm{P}$, which is:
\begin{equation}\label{Hydro}
  \left(
    \begin{array}{c}
      \bm{F}_\mathrm{b} \\
      \bm{L}_\mathrm{b} \\
      \bm{F}_\mathrm{t} \\
      \bm{L}_\mathrm{t} \\
    \end{array}
  \right) = \left(
    \begin{array}{cccc}
      \bm{\xi}_\mathrm{bb}^\mathrm{FU} & \bm{\xi}_\mathrm{bb}^\mathrm{F\omega} & \bm{\xi}_\mathrm{bt}^\mathrm{FU} & \bm{\xi}_\mathrm{bt}^\mathrm{F\omega} \\
      \bm{\xi}_\mathrm{bb}^\mathrm{LU} & \bm{\xi}_\mathrm{bb}^\mathrm{L\omega} & \bm{\xi}_\mathrm{bt}^\mathrm{LU} & \bm{\xi}_\mathrm{bt}^\mathrm{L\omega} \\
      \bm{\xi}_\mathrm{tb}^\mathrm{FU} & \bm{\xi}_\mathrm{tb}^\mathrm{F\omega} & \bm{\xi}_\mathrm{tt}^\mathrm{FU} & \bm{\xi}_\mathrm{tt}^\mathrm{F\omega} \\
      \bm{\xi}_\mathrm{tb}^\mathrm{LU} & \bm{\xi}_\mathrm{tb}^\mathrm{L\omega} & \bm{\xi}_\mathrm{tt}^\mathrm{LU} & \bm{\xi}_\mathrm{tt}^\mathrm{L\omega} \\
    \end{array}
  \right)\cdot
  \left(
    \begin{array}{c}
      \bm{u}_\mathrm{b} \\
      \bm{\omega}_\mathrm{b} \\
      \bm{u}_\mathrm{t}-\bm{u}_0 \\
      \bm{\omega}_\mathrm{t} \\
    \end{array}
  \right)
\end{equation}
where $\bm{u}_0\equiv(\bm{\xi}_\mathrm{tt}^\mathrm{FU} )^{-1}\cdot\bm{F}_\mathrm{act}$. The system is fully determined with two widely used conditions: (i) the free-swimming condition, i.e., $\bm{F}_\mathrm{b}=-\bm{F}_\mathrm{t}\equiv\bm{F}_\mathrm{eff}$ and $\bm{L}_\mathrm{b}=-\bm{L}_\mathrm{t}-(\bm{r}_\mathrm{t}-\bm{r}_\mathrm{b} )\times\bm{F}_\mathrm{t}\equiv-\bm{L}_\mathrm{eff}+\bm{F}_\mathrm{eff}\times(\bm{r}_\mathrm{b}-\bm{r}_\mathrm{t})$; (ii) the rigid body condition, i.e., $\bm{\omega}_\mathrm{b}=\bm{\omega}_\mathrm{t}\equiv\bm{\omega}_0$ and $\bm{u}_\mathrm{t}=\bm{u}_\mathrm{b}+\bm{\omega}_0\times(\bm{r}_\mathrm{t}-\bm{r}_\mathrm{b})$. Eq.\,\ref{Hydro} is then solved for $\bm{F}_\mathrm{eff}$, $\bm{L}_\mathrm{eff}$, $\bm{\omega}_0$, and $\bm{u}_\mathrm{b}$, leading to system evolution $\Delta\bm{\mathrm{R}}=\bm{\mathrm{U}}\Delta t$.

At finite temperature, $\bm{\mathrm{F}}^\mathrm{B}\neq 0$. By integrating Eq.\,\ref{Langevin} over $\Delta t \gg \tau_\mathrm{B}\sim 10^{-6}\mathrm{sec}$ the Brownian time scale but still small so that changes in configuration are not significant, we get two additional terms in system evolution \cite{Ermak1978,Grassia1995,Brady1988}:
\begin{equation}\label{Thermo}
\Delta\bm{\mathrm{R}}=\bm{\mathrm{U}}\Delta t+k_\mathrm{B}T\mathbf{\nabla}\cdot\bm{\mathrm{\xi}}^{-1}\Delta t+\bm{\mathrm{X}}(\Delta t),
\end{equation}
where $\bm{\mathrm{X}}(\Delta t)$ is a random displacement characterized by a multi-variance Gaussian distribution with $\langle\bm{\mathrm{X}}(\Delta t)\rangle=0$ and $\langle\bm{\mathrm{X}}(\Delta t)\bm{\mathrm{X}}(\Delta t)\rangle=2k_\mathrm{B}T\bm{\mathrm{\xi}}^{-1}\Delta t$.

For an entrapped {\it E. coli}, we have $d/R_\mathrm{b}\sim 0.1$ so small that terms in $\bm{\mathrm{\xi}}$ become large, in agreement with experimental observation of suppressed $\bm{\mathrm{X}}(\Delta t)$ \cite{Drescher2011} that we choose to treat as negligible during entrapment. On the other hand, while commonly neglected in previous numerical studies \cite{Giacche2010, Pimponi2016, Eisenstecken2016}, the entropic term $k_\mathrm{B}T\mathbf{\nabla}\cdot\bm{\mathrm{\xi}}^{-1}$ describes the spatial variations in diffusivity, and in the $d\to0$ limit remains a finite constant independent of $d$. Therefore, the evolution equation becomes $\Delta\bm{\mathrm{R}}=\bm{\mathrm{U'}}\Delta t$ with $\bm{\mathrm{U'}}\equiv\left(\bm{u'}_\mathrm{b}, \bm{\omega'}_\mathrm{b}, \bm{u'}_\mathrm{t}, \bm{\omega'}_\mathrm{t}\right)^T=\bm{\mathrm{U}}+k_\mathrm{B}T\mathbf{\nabla}\cdot\bm{\mathrm{\xi}}^{-1}$, where $\bm{\mathrm{U}}$ satisfies Eq.\,\ref{Hydro} and the rigid body condition changes to $\bm{\omega'}_\mathrm{b}=\bm{\omega'}_\mathrm{t}\equiv\bm{\omega}_0$ and $\bm{u'}_\mathrm{t}=\bm{u'}_\mathrm{b}+\bm{\omega}_0\times(\bm{r}_\mathrm{t}-\bm{r}_\mathrm{b})$.

\textit{Numerical simulation --- }For a typical {\it E. coli} in water at room temperature, we estimate that $k_\mathrm{B}T \approx 4\times 10^{-21} \mathrm{N \cdot m}$, $\mu\approx 10^{-3} \mathrm{N \cdot sec/m^2}$, $|\bm{F}_\mathrm{act}| \approx 2\times 10^{-13} \mathrm{N}$, $R_\mathrm{b} \approx 1 \mathrm{\mu m}$, $l \approx 5 \mathrm{\mu m}$, and $R_\mathrm{t} \approx 0.4 \mathrm{\mu m}$. Using $1 \mathrm{\mu m}$ and $1 \mathrm{sec}$ as the unit of length and time respectively and setting $\mu=1$ for the unit of force, in our study such a bacterium can be characterized by $\{l=5,R_\mathrm{b}=1,R_\mathrm{t}=0.4,|\bm{F}_\mathrm{act}|=200,k_\mathrm{B}T=4\}$. In Fig.\,\ref{Fig1}a we show its dynamics moving towards the plane with initial configuration $\{d=20,\theta=75^\circ\}$, simulated assuming only hydrodynamic interactions without steric interactions. The key in this simulation is getting $\bm{\xi}$ for each configuration, which is constructed following the Stokesian dynamics simulation \cite{Brady1988,Swan2007} in a two-step procedure. Specifically, we first model the grand mobility tensor in the far-field limit through $\bm{M}_{\infty}=\bm{M}_{0}+\bm{\hat{M}}$. Here $\bm{M}_{0}$ is the analytic far-field hydrodynamic interaction without plane \cite{Jeffrey1984}, and $\bm{\hat{M}}$ corresponds to the plane contribution which is analytically available through the method of images \cite{Swan2007}. In the second step, $\bm{\xi}$ is obtained through $\bm{\xi}=\bm{M}_{\infty}^{-1}+\bm{\xi}_\mathrm{b}$, where $\bm{\xi}_\mathrm{b}$ characterizes the lubrication between body-bead and the plane. Once $\bm{\xi}$ is obtained, we can solve Eq.\,\ref{Hydro} for $\bm{F}_\mathrm{eff}$, $\bm{L}_\mathrm{eff}$, $\bm{\omega}_0$, and $\bm{u'}_\mathrm{b}$.

\begin{figure}[t]
\includegraphics[width=.5\textwidth]{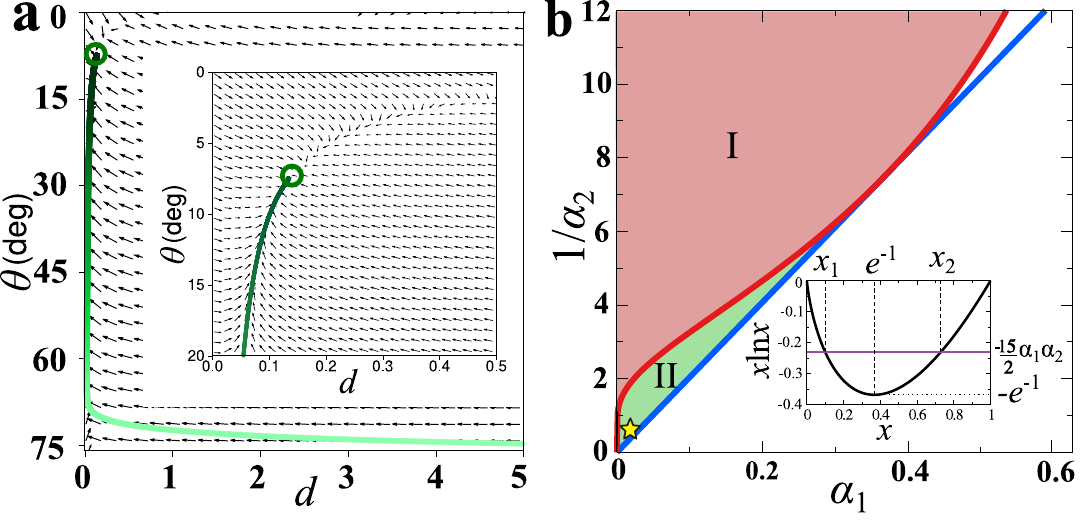}
\caption{(a) In the phase diagram defined by $d$ and $\theta$, bacterial trajectory (green line) approaches to a stable fixed point (green circle). (b) The parameter space defined by $\alpha_1$ and $\alpha_2$ can be divided into three regions: region I (red) above the red line $\alpha_2=-\frac{2}{15}\ln{\alpha_1}$, region II (green) between the red line and blue line $\frac{15}{2}\alpha_1\alpha_2=e^{-1}$ with $\alpha_1\leq e^{-1}$, and the region outside the above two regions. A typical {\it E. coli} at room temperature is characterized by the yellow star in region II. The inset in (b) shows $x\ln{x}$ as a function of $x$.}\label{Fig2}
\end{figure}

As illustrated by the temporal evolution of $d$ and $\theta$ in Fig.\,\ref{Fig1}b, Our numerical results reproduce the experimentally observed three-stage dynamics \cite{Bianchi2017}: the initial approach where $d$ drops very quickly with an almost constant $\theta$; the reorientation stage where $\theta$ decreases very quickly right after $d$ falls below the size of body-bead; and the steady swimming stage where both $\dot{d}$ and $\dot{\theta}$ gradually decay to zero characterizing a stable entrapment. Such dynamics can also be illustrated in the phase diagram defined by $d$ and $\theta$ (Fig.\,\ref{Fig2}a), where the steady swimming corresponds to a stable fixed point with $\theta>0$ (a “nose-down” configuration), in agreement with previous experimental observations. A quantitative study of the reorientation stage shows that $\tan{\theta}$ decays exponentially for $t>t_0$ with $t_0$ the time when $d$ falls below the size of body-bead (Fig.\,S1 in SI), in good agreement with experimental observations (Fig.\,3d in \cite{Bianchi2017}). While this exponential decay was attributed to steric interactions in \cite{Bianchi2017}, in our simulation it is obtained assuming only hydrodynamic interactions.

\textit{Analytic solution --- }To understand these results, we simplify the problem with two ideal approximations for analytic solutions. Since $l \gg R_\mathrm{t}$, our first approximation assumes that the tails are not hydrodynamically coupled with the body or the plane. Specifically, we have body-tail coupling $\bm{\xi}_\mathrm{bt}=\bm{\xi}_\mathrm{tb}=0$, and tail self term $\bm{\xi}_\mathrm{tt}$ a constant. Therefore, the only configurational dependent term in $\bm{\mathrm{\xi}}$ is $\bm{\xi}_\mathrm{bb}$, which is an analytical function of single parameter $d⁄R_\mathrm{b}$ in the $d\to 0$ limit \cite{Jeffrey1984}. Since $d/R_\mathrm{b}$ is small during entrapment, our second approximation uses this analytical $\bm{\xi}_\mathrm{bb}$ function for all $d$ of interest. These two ideal approximations lead to a simplified $k_\mathrm{B}T\mathbf{\nabla}\cdot\bm{\mathrm{\xi}}^{-1}$ with only one non-zero component: A constant translational velocity $\bm{v}=\frac{k_\mathrm{B}T}{6\pi\mu R_\mathrm{b}^2}\bm{\hat{x}}$ for bacterial body moving away from the plane (SI, Sec.\,A).

To obtain fixed points in the phase diagram defined by $d$ and $\theta$, we insert $\dot{d}=\dot{\theta}=0$ to Eq.\,\ref{Hydro}, which gives:
\begin{eqnarray}
(\bm{\xi}_\mathrm{bb}^\mathrm{LU})_{zy}*(\bm{u'}_\mathrm{b}\cdot\bm{\hat{y}})&=&(\bm{\xi}_\mathrm{tt}^\mathrm{FU})_{yy}*(\bm{u'}_\mathrm{b}\cdot\bm{\hat{y}})*l*\sin{\theta} \label{Torque} \\
|\bm{F}_\mathrm{act}|*\sin{\theta}&=&(\bm{\xi}_\mathrm{bb}^\mathrm{FU})_{xx}*(\bm{v}\cdot\bm{\hat{x}}) \label{Force}
\end{eqnarray}
Here Eq.\,\ref{Torque} is equivalent to the fifth equation in \cite{Sipos2015}, which characterizes the torque balance on the body-bead in $\bm{\hat{z}}$, between the boundary-induced torque due to bacterial body translation $\bm{u'}_\mathrm{b}\cdot\bm{\hat{y}}$ along the plane (LHS) and the torque arises from the friction against bacterial tail translation (RHS). Eq.\,\ref{Force} characterizes the force balance on the body-bead in $\bm{\hat{x}}$, between self-propulsion (LHS) and the entropic effect we introduced (RHS). The coefficients in Eq.\,\ref{Torque} and Eq.\,\ref{Force} are available from lubrication theory as $(\bm{\xi}_\mathrm{bb}^\mathrm{LU})_{zy}=-6\pi\mu R_\mathrm{b}^2*\frac{2}{15}\ln{\frac{d}{R_\mathrm{b}}}$, $(\bm{\xi}_\mathrm{tt}^\mathrm{FU})_{yy}=6\pi\mu R_\mathrm{t}$, $(\bm{\xi}_\mathrm{bb}^\mathrm{FU})_{xx}=\frac{6\pi\mu R_\mathrm{b}^2}{d}$.

When $\bm{u'}_\mathrm{b}\cdot\bm{\hat{y}}=0$, a trivial solution $\{\theta_0=\frac{\pi}{2},d_0=\frac{k_\mathrm{B}T}{|\bm{F}_\mathrm{act}|}\}$ exists, where the bacterium points straightly towards the plane. For $\bm{u'}_\mathrm{b}\cdot\bm{\hat{y}}\neq 0$, Eq.\,\ref{Torque} and Eq.\,\ref{Force} can be further reduced in terms of two dimensionless parameters $\alpha_1\equiv \frac{k_\mathrm{B}T}{|\bm{F}_\mathrm{act}|R_\mathrm{b}}$ and $\alpha_2\equiv \frac{R_\mathrm{t}l}{R_\mathrm{b}^2}$:
\begin{eqnarray}
\frac{d}{R_\mathrm{b}}\ln{\frac{d}{R_\mathrm{b}}}&=&-\frac{15\alpha_1 \alpha_2}{2} \label{External}\\
\ln{\frac{d}{R_\mathrm{b}}}&=&-\frac{15\alpha_2 \sin{\theta}}{2} \label{Intrinsic}
\end{eqnarray}
where the prefactor -15/2 arises from translation-rotation coupling $(\bm{\xi}_\mathrm{bb}^\mathrm{LU})_{zy}$.

Dictated by Eq.\,\ref{External} and Eq.\,\ref{Intrinsic}, two curves, $\frac{15}{2}\alpha_1\alpha_2=e^{-1}$ and $\alpha_2=-\frac{2}{15}\ln{\alpha_1}$, become important in the parameter space defined by $\alpha_1$ and $\alpha_2$ (the blue line and the red line, respectively, in Fig.\,\ref{Fig2}b). The first curve, $\frac{15}{2}\alpha_1\alpha_2=e^{-1}$, arises from the fact that $\frac{d}{R_\mathrm{b}}\ln{\frac{d}{R_\mathrm{b}}}\geq-e^{-1}$ for $d>0$ where the equality happens only at $\frac{d}{R_\mathrm{b}}=e^{-1}$ (Fig.\,\ref{Fig2}b inset). Thus, nontrivial fixed points ($\left\{d_1,\theta_1\right\}$ and $\left\{d_2,\theta_2\right\}$) only exist when $0\geq -\frac{15}{2}\alpha_1\alpha_2\geq -e^{-1}$, with $0\leq \frac{d_1}{R_\mathrm{b}}\leq e^{-1}\leq \frac{d_2}{R_\mathrm{b}}\leq 1$. The second curve, $\alpha_2=-\frac{2}{15}\ln{\alpha_1}$, is obtained by assuming $\sin{\theta}=1$, the largest possible value for $\sin{\theta}$. Our analysis shows that (Sec.\,C in SI), for parameter choice $\{\alpha_1,\alpha_2\}$ above the red line in Fig.\,\ref{Fig2}b (region I), only the nontrivial fixed point $\left\{d_2,\theta_2\right\}$ with $\frac{d_2}{R_\mathrm{b}}>e^{-1}$ exists. For parameter choice $\{\alpha_1,\alpha_2\}$ between the red line and blue line with $\alpha_1\leq e^{-1}$ in Fig.\,\ref{Fig2}b (region II), both nontrivial fixed points exist. No nontrivial fixed point exists for parameter choice $\{\alpha_1,\alpha_2\}$ outside region I and II.

Furthermore, for various choices of $\{\alpha_1,\alpha_2\}$ in region I and II, we numerically studied the neighborhood of each fixed point in the phase diagram defined by $d$ and $\theta$. Our results for all cases (SI, Sec.\,C) show that $\{d_1,\theta_1\}$ with $\frac{d_1}{R_\mathrm{b}}<e^{-1}$ is always stable, while $\{d_2,\theta_2\}$ with $\frac{d_2}{R_\mathrm{b}}>e^{-1}$ is always a saddle point. Since physically observed entrapment is associated only with a stable fixed point which only exists for $\{\alpha_1,\alpha_2\}$ in region II (Fig.\,\ref{Fig2}b), this region defines the entrapment zone: No physical entrapment can be observed for $\{\alpha_1,\alpha_2\}$ outside the zone.

For a typical {\it E. coli} at room temperature, we have $\alpha_1=0.02$ and $\alpha_2=2$, which falls in the entrapment zone (Fig.\,\ref{Fig2}b). For this set of $\{\alpha_1,\alpha_2\}$, we predict the stable fixed point at $d^*=0.17 \mathrm{\mu m}$ from Eq.\,\ref{External} and then $\theta^*=7^\circ$ from Eq.\,\ref{Intrinsic}. Using our simulation that includes hydrodynamic coupling of the tails, we get $d^*=0.13 \mathrm{\mu m}$ and $\theta^*=8^\circ$. Both our analytic and simulation results are in good agreement with experimental observations \cite{Bianchi2017,Li2008}.

\begin{figure}[t]
\includegraphics[width=.48\textwidth]{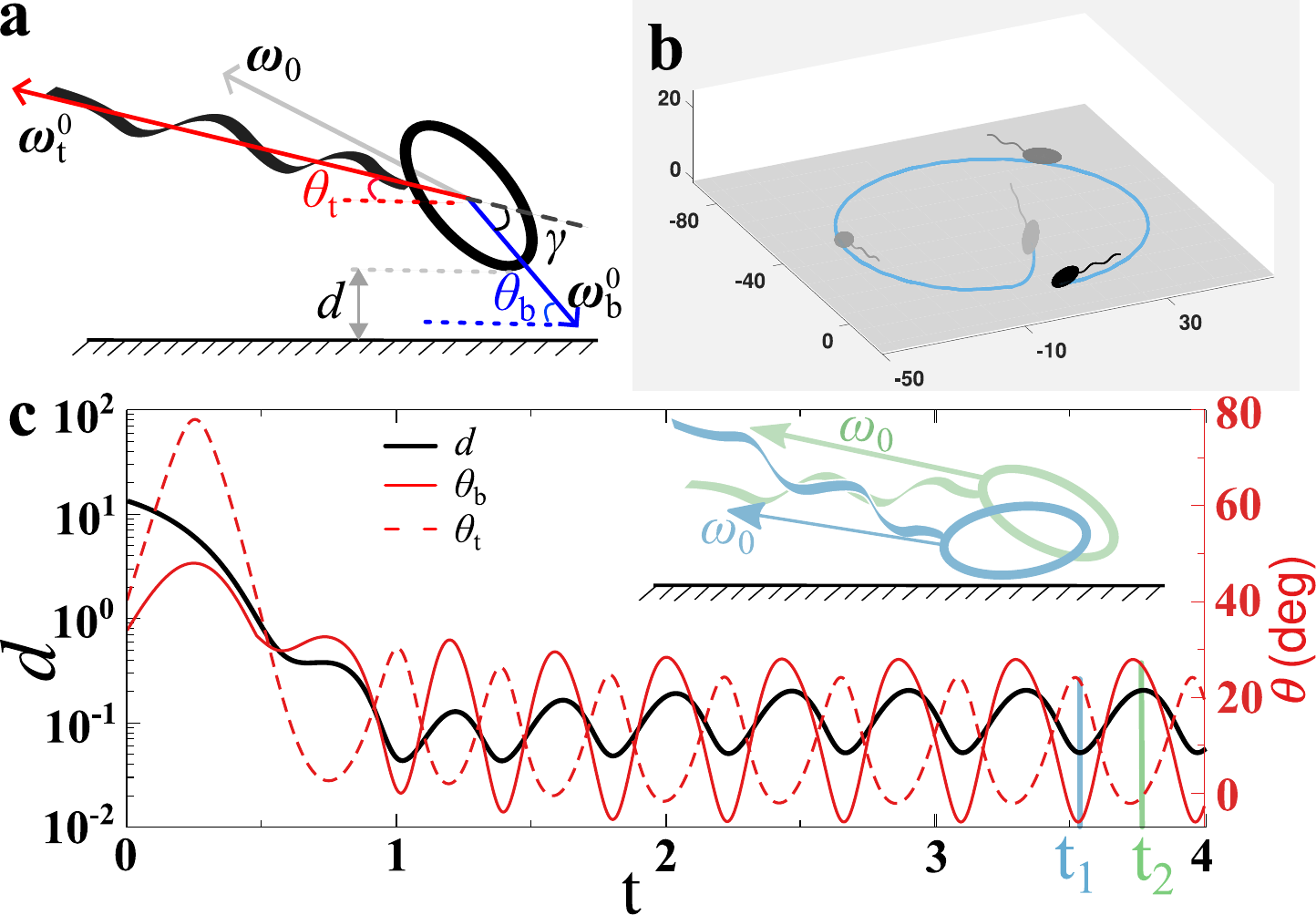}
\caption{(a) Our model of a wobbling bacterium explicitly considers $\bm{\omega}_\mathrm{b}^0$ and $\bm{\omega}_\mathrm{t}^0$. (b) Trajectory of a wobbling bacterium near a plane. (c) The temporal evolutions of $d$ (solid black), $\theta_\mathrm{b}$ (solid red), and $\theta_\mathrm{t}$ (dash red). Blue (green) bar indicates time $t_1$ ($t_2$) when $\theta_\mathrm{b}$ is smallest (largest) during entrapment, with the corresponding configurations shown in the inset using actual $\theta_\mathrm{b}$ and $\theta_\mathrm{t}$, and exaggerated $d$.}\label{Fig3}
\end{figure}

\textit{Bacterial wobbling --- }In previous sections, we obtained analytic predictions for non-wobbling bacteria, where the body-tail connection is treated as rigid without considering the self-spinning of either the bacterial body or the flagellar bundle, i.e., $\bm{\omega}_\mathrm{b}=\bm{\omega}_\mathrm{t}=\bm{\omega}_0$. In real world, the self-spinning of the two parts, $\bm{\omega}_\mathrm{b}^0$ and $\bm{\omega}_\mathrm{t}^0$, respectively, are generally not collinear (Fig.\,\ref{Fig3}a). For free-swimming bacteria the nonzero angle $\gamma$ formed by the two vectors leads to a center of mass rotation $\bm{\omega}_0$ generally not aligned with its translation and therefore, bacterial wobbling. To account for the wobbling, we generalize the non-wobbling bacterial model by explicit consideration of $\bm{\omega}_\mathrm{b}^0$ and $\bm{\omega}_\mathrm{t}^0$, and replace the rigid connection by a universal joint through relation $R_\mathrm{b}^3|\bm{\omega}_\mathrm{b}^0|\cos{\gamma}=R_\mathrm{t}^3|\bm{\omega}_\mathrm{t}^0|$ \cite{Kamdar2022, Shimogonya2016} where $|\bm{\omega}_\mathrm{t}^0|=\mathrm{100 Hz}$ fixed to match the experiments \cite{Darnton2007} (SI, Sec.\,D). The overall rotations for bacterial body and flagellar bundle are $\bm{\omega}_\mathrm{b}=\bm{\omega}_\mathrm{b}^0+\bm{\omega}_0$ and $\bm{\omega}_\mathrm{t}=\bm{\omega}_\mathrm{t}^0+\bm{\omega}_0$, respectively.

With this generalisation, we numerically study the dynamics of wobbling bacteria near a plane, where each configuration is now determined by $d$ and two distinct pitch angles, $\theta_\mathrm{b}$ and $\theta_\mathrm{t}$ (Fig.\,\ref{Fig3}a). For illustration purposes, in Fig.\,\ref{Fig3} we highlight the self-spinning by drawing an ellipsoid and a helix in place of the body-sphere and tail-sphere used in our actual simulations. Our results show that bacteria can be trapped in a clockwise circular trajectory when viewed from above (Fig.\,\ref{Fig3}b). The radii of the circles are $10\sim 200 \mathrm{\mu m}$, which increase with $d$ and $\alpha_1$ (Fig.\,S4 in SI), all of which are in agreement with earlier experiments \cite{Lauga2006,Cao2022,Li2008,Maeda1976}. While for technical reasons most experiments focus on $\theta_\mathrm{b}$ with scarce data available for $\theta_\mathrm{t}$, our simulation has the advantage to capture both. For a typical bacterium ($l=5$, $R_\mathrm{b}=1$, $R_\mathrm{t}=0.4$) with $\gamma=30^\circ$ at room temperature, in Fig.\,\ref{Fig3}c we show our numerical results for the temporal evolution of $d$, $\theta_\mathrm{b}$, and $\theta_\mathrm{t}$. During the entrapment stage, all these three variables are periodically oscillating with the same frequency determined by $\bm{\omega}_0$, where $d$ and $\theta_\mathrm{b}$ are almost in phase while $\theta_\mathrm{t}$ has nearly an opposite phase (Fig.\,\ref{Fig3}c). In the inset, we show the bacterial configurations at $t_1$ ($t_2$) with smallest (largest) $\theta_\mathrm{b}$ denoted as $\theta_\mathrm{min}$ ($\theta_\mathrm{max}$). The experimentally recorded average pitch angle $\Bar{\theta}$ and wobbling angle $\theta_w$ \cite{Bianchi2017} can be obtained through $\Bar{\theta}\equiv (\theta_\mathrm{max}+\theta_\mathrm{min})/2$ and $\theta_w\equiv (\theta_\mathrm{max}-\theta_\mathrm{min})/2$. In a similar fashion, we define $\Bar{d}\equiv (d_\mathrm{max}+d_\mathrm{min})/2$.

\begin{figure}[t]
\includegraphics[width=.48\textwidth]{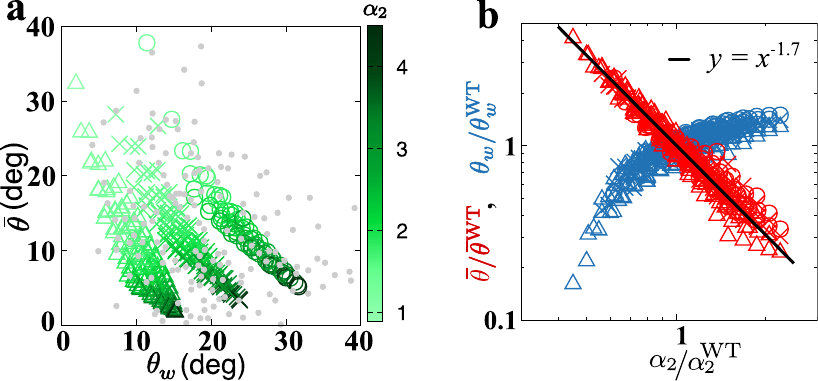}
\caption{Simulation data with $\gamma=20^\circ$ (triangles), $\gamma=30^\circ$ (crosses), and $\gamma=40^\circ$ (circles) yield the scatter plot $\{\Bar{\theta},\theta_w\}$ (green) in (a), normalized $\Bar{\theta}$ (red) and $\theta_w$ (blue) as functions of $\alpha_2$ in (b). Results from \cite{Bianchi2017} are shown as grey dots in (a).}\label{Fig4}
\end{figure}

A variety of bacteria characterized by $\{l,R_\mathrm{b},R_\mathrm{t},\gamma\}$ are then simulated at room temperature, where for $\gamma$ we sampled $20^\circ$, $30^\circ$, and $40^\circ$, and for each of the other three parameters we limit the variations to at most $20\%$ from the corresponding value of a typical {\it E. coli}. The scatter plot of all $\{\Bar{\theta},\theta_w\}$ obtained from the entrapment stage show an anticorrelation spreading broadly along the $\Bar{\theta}=\theta_w$ direction (Fig.\,\ref{Fig4}a), in quantitative agreement with previous experiment (Fig.\,4 in \cite{Bianchi2017}).

Interesting results emerge when sorting out our data by $\gamma$. For each specific $\gamma$ we observe an anticorrelation with a much narrower spread, which is regulated by $\alpha_2$ in a quantitatively similar fashion. Specifically, for each $\gamma$ we plot $\Bar{\theta}$ ($\theta_w$) as a function of $\alpha_2$, normalized by $\Bar{\theta}^\mathrm{WT}$ ($\theta_w^\mathrm{WT}$) obtained from a typical bacterium with that particular $\gamma$. Our data from three distinct $\gamma$ collapse, indicating existence of two master curves for $\Bar{\theta}/\Bar{\theta}^\mathrm{WT}$ and $\theta_w/\theta_w^\mathrm{WT}$ respectively (Fig.\,\ref{Fig4}b).

Our observation that the anticorrelation is dictated mostly by variation in $\alpha_2$ can be explained by Eq.\,\ref{External} and Eq.\,\ref{Intrinsic}, which are still valid for wobbling bacteria if we replace $\{d, \theta\}$ in the equations by $\{\Bar{d}, \Bar{\theta}\}$ observed in data (Fig.\,S5 in SI). Since we are in the parameter range that changes in $\ln{(\Bar{d}/R_\mathrm{b})}$ are much less significant than changes in $\Bar{d}/R_\mathrm{b}$, for an estimate we can neglect changes in $\ln{(\Bar{d}/R_\mathrm{b})}$, which leads to $\Bar{d}/R_\mathrm{b}\sim \alpha_1\alpha_2$ according to Eq.\,\ref{External} and $\sin{\Bar{\theta}}\approx \Bar{\theta}\sim \alpha_2^{-1}$ independent of $\alpha_1$ according to Eq.\,\ref{Intrinsic}. This estimate correctly captures the positive (negative) correlation between $\Bar{d}/R_\mathrm{b}$ ($\Bar{\theta}$) and $\alpha_2$, in qualitative agreement with numerical fit of our data, which gives $\Bar{d}/R_\mathrm{b}\sim \alpha_2^{1.5}$ (Fig.\,S6a), $\ln{(\Bar{d}/R_\mathrm{b})}\sim \alpha_2^{-0.8}$ (Fig.\,S6b), and $\Bar{\theta}\sim \alpha_2^{-1.7}$ (Fig.\,\ref{Fig4}b). As it is experimentally established that wobbling can be significantly suppressed by nearby boundary surfaces \cite{Kamdar2022,Vizsnyiczai2020}, the positive correlation between $\Bar{d}/R_\mathrm{b}$ and $\alpha_2$ leads to positive correlation between $\theta_w$ and $\alpha_2$ (more suppressed wobbling at smaller $\alpha_2$), and thus the anticorrelation between $\Bar{\theta}$ and $\theta_w$.

\textit{Discussion --- }In this study we ignore stochastic effects $\bm{\mathrm{X}}(\Delta t)$ and bacterial tumbling. At a cost of making trapped bacteria incapable to escape, this simplification helps us focus on the entrapment stage and highlight the importance of the entropic effect $k_\mathrm{B}T\mathbf{\nabla}\cdot\bm{\mathrm{\xi}}^{-1}$. While surface escape at large time scales is out of scope for current work, it is interesting for a future study to investigate how variations in $\alpha_1$ and $\alpha_2$ influence stability of fixed points against stochastic fluctuations, leading to parameter dependence of surface residence time.

For the entrapment stage, we show that the self propulsion for the “nose-down” configuration needs to be balanced by a cell-plane repulsion, in sharp contrast to the earlier notion that surface entrapment arises from an effective attraction \cite{Frymier1995}. At room temperature and $d\approx 0.1 \mathrm{\mu m}$, an entropic repulsion arises naturally from the drastic spatial variation of near field hydrodynamic interactions. By including this entropic term, we demonstrate that hydrodynamic interactions alone explain existing observations, with even a simplified model that considers bacterial body (flagellar bundle) as a sphere with the size equal to its hydrodynamic radius. More importantly, for future experiments to verify if other mechanisms such as steric interactions or stochastic effect $\bm{\mathrm{X}}(\Delta t)$ also contribute, our model has provided the following predictions: (i) The entrapped configuration $\{d,\theta\}$ is dictated by two dimensionless parameters $\alpha_1\equiv\frac{k_\mathrm{B}T}{|\bm{F}_\mathrm{act}|R_\mathrm{b}}$ and $\alpha_2\equiv \frac{R_\mathrm{t}l}{R_\mathrm{b}^2}$, through two analytic relations, i.e., Eq.\,\ref{External} and Eq.\,\ref{Intrinsic}. Extra care is needed for studies at different temperatures, as self propulsion can triple with a slight increase of $k_\mathrm{B}T$ from $20^\circ \mathrm{C}$ to $40^\circ \mathrm{C}$ \cite{Maeda1976}. (ii) About the flagellar bundle, the orientation $\theta_\mathrm{t}$ has an opposite phase with respect to $d$ and $\theta_\mathrm{b}$; and the variation in angle $\gamma$ is the main factor underlying the broad spread of observed $\{\Bar{\theta},\theta_w\}$ data along the $\Bar{\theta}=\theta_w$ direction. (iii) There exists an entrapment zone within the range of $0<\alpha_1<e^{-1}$ and $\alpha_2>2/15$.

Three implications follow from (iii). First, the existence of an upper limit in $\alpha_1$ demonstrates that entrapment is a special feature for active systems, and no entrapment is allowed for passive systems (zero activity and thus $\alpha_1=\infty$) at any temperature. Second, entrapment is not available for any active swimmers with intrinsic shape factor $\alpha_2\equiv \frac{R_\mathrm{t}l}{R_\mathrm{b}^2}<2/15$, which provides a guideline for controlling biological and engineering active swimmers near surfaces. Third, while entrapment only exists at low temperatures that satisfy $\alpha_1\equiv\frac{k_\mathrm{B}T}{|\bm{F}_\mathrm{act}|R_\mathrm{b}}<e^{-1}$, at the lowest temperature possible, i.e., the absolute zero temperature, our solution becomes singular with surface distance $\lim\limits_{k_\mathrm{B}T\to0}d_1\to0$, where all theoretical studies treating the solvent as a continuum through hydrodynamics break down. This explains why negligence of $k_\mathrm{B}T\mathbf{\nabla}\cdot\bm{\mathrm{\xi}}^{-1}$ in previous studies of hydrodynamic interactions, which is equivalent to setting the temperature to zero, cannot reproduce the entrapment correctly. Instead, a finite temperature is essential in achieving the physical entrapment, or in other words, sticky bacteria are hot.
\parskip = 5pt plus 1pt

\parskip = 5pt plus 1pt

We thank X. Cheng, L. S. Luo, M. Stynes, Y. L. Wu, J. H. Yuan, and H. P. Zhang for helpful discussions. This work is supported by NSFC No. 11974038 and No. U2230402. We also acknowledge the computational support from the Beijing Computational Science Research Center.

\bibliography{EntrapmentPRL}

\end{document}